\documentclass[superscriptaddress,twocolumn,showpacs,preprintnumbers,prl]{revtex4}
\usepackage{graphicx,color}
\usepackage{times}
\usepackage{epsfig}
\begin{document}

\def\salto{\vskip 1cm}
\def\lag{\langle}
\def\rag{\rangle}

\newcommand{\LANL} {Theoretical Division T-11, Los Alamos National Laboratory, Los Alamos, NM 87545, USA}
\newcommand{\UCSB} {Microsoft Project Q, The University of California, Santa Barbara, CA 93106, USA}
\newcommand{\UM} {Condensed Matter Theory Center, Department of Physics, The University of Maryland, College Park, MD 
20742, USA}

\title{Topological Confinement and Superconductivity}

\author{K. A. Al-Hassanieh }   \affiliation {\LANL}
\author{C. D. Batista}                \affiliation {\LANL} 
\author{P. Sengupta}                \affiliation {\LANL}
\author{A. E. Feiguin }              \affiliation {\UM} \affiliation{\UCSB}

\begin{abstract}
We derive a Kondo Lattice model with a correlated conduction band from a two-band Hubbard Hamiltonian. This 
mapping allows us to describe the emergence of a robust pairing mechanism in a model that only contains 
repulsive interactions. The mechanism is 
due to topological confinement and results from the interplay between antiferromagnetism and delocalization. 
By using Density-Matrix-Renormalization-Group (DMRG) 
we demonstrate that this mechanism leads to dominant superconducting correlations in a 1D-system. 
\end{abstract}


\maketitle

The origin of unconventional superconductivity remains as one the most important open problems
of condensed matter physics. Physicists do not agree on the mechanism that pairs the electrons to
form the superconducting  condensate. To a 
very good approximation, electrons only interact via the repulsive Coulomb interaction.
Consequently, as it was pointed out recently by Anderson \cite{Anderson07}, the crucial question is:
``How can this repulsion between electrons be eliminated in favor of electron pair binding?'' 
The problem becomes even more puzzling if we consider that 
the Coulomb interaction is bigger than the bandwidth for most of the unconventional superconductors. 
Then, the first  challenge is to demonstrate that a ``pairing force'' can exist in
model Hamiltonians that only contain strongly repulsive interactions. 
Although different pairing
mechanisms have been proposed over the last twenty years, it is not always clear if 
those mechanisms actually work or if they are robust under the presence of long range Coulomb interactions.
This is mainly due to the lack of controlled approximations for solving models of strongly interacting
electrons in 2D or 3D. 

Another aspect that is quite ubiquitous in unconventional superconductors is the proximity
of the superconducting state to an antiferromagnetic (AFM) phase. This observation suggests that
antiferromagnetic correlations are related to the pairing mechanism. However, although several
``magnetic'' pairing mechanisms have been proposed \cite{Dagotto94}, it is still unclear how the interplay between
AFM correlations and itineracy leads to a ``glue'' that is strong enough to hold the two electrons
together. It is the purpose of this Letter to show how a robust pairing mechanism emerges out of this interplay
and to demonstrate that it leads to dominant superconducting correlations in a two-band Hubbard chain.
Moreover, we will see that the pairing is still robust in the proximity of the AFM region, i.e., when 
there are large AFM fluctuations but no AFM order.    
What makes the mechanism robust is the fact that it is driven by confinement of topological defects (solitons)
that are attached to each carrier (holes). In particular, as we will see  
below, this implies that the binding energy and the size of the pair are determined by different energy scales.

We start by deriving an extended Kondo lattice (KL) chain with a correlated ($t-J$) conduction band
as the low-energy effective model ${\tilde H}$ of the original two-band Hubbard Hamiltonian $H$.
The correlated nature of the conduction band is the main difference with the standard KL chain that
was extensively studied in previous works \cite{Tsunetsugu97}.  
With a simple analytical treatment of the fully anisotropic (Ising-like) limit of ${\tilde H}$,
we show the origin of the two-hole bound state and the corresponding binding energy in the dilute limit.
Our DMRG calculations allow us to extend these results to the fully isotropic (Heisenberg) limit 
and to finite hole concentrations. Moreover, we show that the superconducting pair-pair correlations
are the dominant ones over an extended and relevant region of the quantum phase diagram. Interestingly enough,
the pairing remains robust in the absence of long-range AFM order (fully isotropic limit) indicating that a long enough
AFM correlation length is sufficient for stabilizing the pairing mechanism that we discuss below.

We will consider the following two-band Hubbard chain:
\begin{eqnarray}
H &=& \!\!\!\!\sum_{j;\sigma;\eta} \!\! (e_{\eta}-\mu) n_{j \sigma\eta} +
t_{\eta \eta} (c^{\dagger}_{j+1 \sigma\eta} c^{\;}_{j \sigma\eta} + c^{\dagger}_{j \sigma\eta} c^{\;}_{j+1 \sigma\eta})
\nonumber \\
&+& \!\!\!\! \!\sum_{j;\sigma} t_{ul}
(c^{\dagger}_{j \sigma u} c^{\;}_{j \sigma l} + c^{\dagger}_{j \sigma l } c^{\;}_{j \sigma u}) 
+ \! \!\sum_{j;\eta}  U_{\eta} n_{j \uparrow\eta} n_{j \downarrow\eta}, 
\end{eqnarray} 
where $1 \leq j \leq L$, $L$ is even, $L+1 \equiv 1$ [periodic boundary conditions (PBC)], { $\eta=\{l,u\}$ 
denotes the lower and upper bands, and $\sigma=\{\uparrow,\downarrow\}$.}
The diagonal energies are $e_{u}=\Delta_{ul}/2$ and $e_{l}=-\Delta_{ul}/2$ with $\Delta_{ul}>0$, and the density
operators are $n_{j \sigma\eta}=c^{\dagger}_{j \sigma \eta} c^{\;}_{j \sigma \eta}$.

From now on, we will assume that the mean number of electrons per unit cell is $1 \leq n \leq 2$.
For $t_{\eta \eta'}=0$ and $U_u > \Delta_{ul}$, the ground state subspace, ${\cal S}$, consists of states
containing one electron per site in the lower band (only spin remains as a degree of freedom).
In contrast, the sites of the upper band can be empty or singly occupied. 
In the strong coupling limit, $U_u, U_l, \Delta_{ul} \gg t_{\eta \eta'}$ and $U_l-\Delta_{ul} \gg t_{ul}$, 
the low-energy spectrum of $H$ is described by the effective Hamiltonian, ${\tilde H}$, 
that acts on the subspace ${\cal S}$ and
results from applying degenerate perturbation theory to second order in the hopping terms:
\begin{eqnarray}
{\tilde H} &=& t_{uu} \sum_{j;\sigma} 
({\bar c}^{\dagger}_{j+1 \sigma u} {\bar c}^{\;}_{j \sigma u} + {\bar c}^{\dagger}_{j \sigma u } {\bar c}^{\;}_{j+1 \sigma u})
+ \sum_{j;\sigma}  (e_{u}-{\tilde \mu}) {\bar n}_{j \sigma u} 
\nonumber \\
&+& 
\sum_{j\nu \eta} J^{\nu}_{\eta} S^{\nu}_{j\eta} S^{\nu}_{j+1\eta} +
\sum_{j\nu}  J^{\nu}_K S^{\nu}_{ju} S^{\nu}_{jl},
\end{eqnarray} 
where ${\bar c}^{\dagger}_{j \sigma u}= c^{\dagger}_{j \sigma u} (1-n_{j {\bar \sigma} u})$ (constraint
of no double-occupancy), $\nu=\{x,y,z\}$,
$J^{\nu}_{\eta}=4t^2_{\eta\eta}/U_{\eta}$, and $J^{\nu}_K=2t^2_{ul}/(U_l-\Delta_{ul})+2t^2_{ul}/(U_u+\Delta_{ul})$.
Although the exchange interactions are isotropic ($J^{\nu}_{\eta}=J_{\eta}$ and $J^{\nu}_K=J_K$), we split the Heisenberg
terms for reasons that will become clear later.  Note that we have neglected the attractive, 
$-\frac{J_u}{4} {\bar n}_j {\bar n}_{j+1}$, and the correlated hopping terms that also appear to
second order in $t_{uu}$ to keep ${\tilde H}$ simple and because
they are not relevant for the pairing derived below. A more extensive study including the effect of these
terms will be presented elsewhere \cite{Khaled}.
To simplify the notation, we will introduce $t=t_{uu}$, $J = J^{z}_{u}$, 
$\alpha J = J^{x}_{u}  = J^{y}_{u}$, $J_H = J^{z}_{l}$, and $ \beta J_H = J^{x}_{l}  = J^{y}_{l}$.  
Where $0\le \alpha,\beta \le 1$ determine the exchange anisotropies. ${\tilde H}$ is an extension of the Kondo Lattice
chain: the localized spins of the lower band are coupled via a Kondo interaction, $J_K$, to the itinerant electrons  
of the upper band which are described by a $t-J$ model. In contrast to the usual Kondo Lattice, the itinerant electrons
are strongly correlated and the local spins are AFM coupled via $J_H$. 

In the following, we set $t=1$ as the  energy scale and use $J = 0.4$, $J_H = 0.5$. These values correspond
for instance to $U_u=10$, $U_l=16$ and $t_{ll}=\sqrt{2}$. We will start by assuming an 
Ising-like ($\beta = 0$) coupling between the localized spins
to  stabilize long-range antiferromagnetic order at $T=0$. The fully isotropic case $\alpha=\beta=1$ will be considered 
in the second part of the manuscript, so $\beta=0$ unless its value is explicitly specified.
Below we will drop the band index in the definition
of the different correlation functions because they are always applied to the conduction band.  We will only compute the 
ground state properties. 

We use the DMRG method \cite{White92,Hallberg06} to study systems up to 100 unit cells at $T=0$.  
The PBC increase substantially the computational effort.  In the finite-system step, we keep up to $M = 1400$ states per block and perform up to 12 sweeps.  The weight 
of the discarded states is kept to the order $10^{-6}-10^{-10}$ for $\beta=0$, and $10^{-5}$ for $\beta=1$
\cite{note1}.

\begin{figure}[t]
\includegraphics[angle=0,width=8cm]{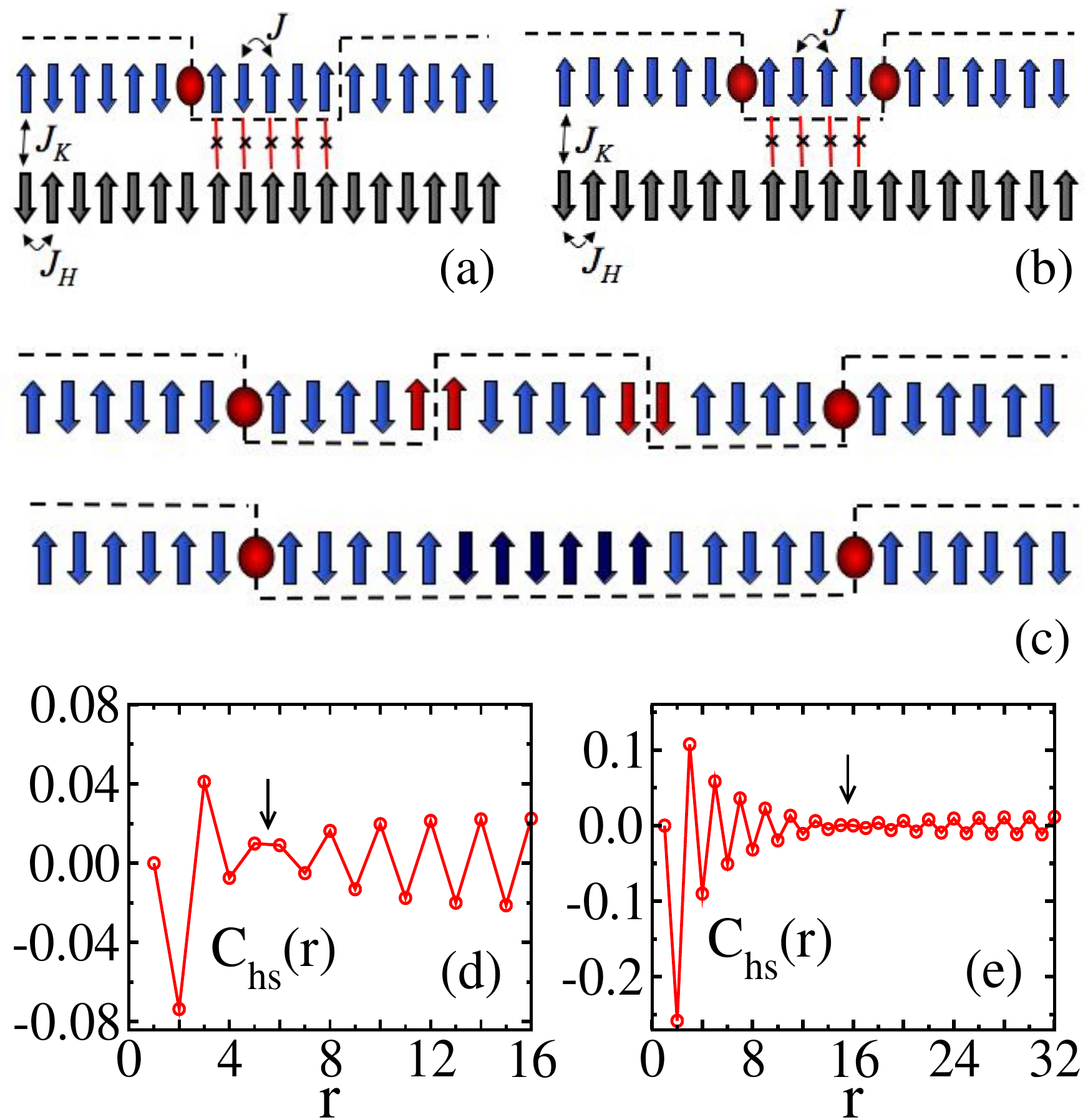}
\vspace*{-0.3cm}
\caption{(Color online) (a) Scheme of the single-hole quasiparticle.  
The holon and spinon carry solitons with opposite topological charges as indicated by the dashed line.   The 
itinerant and localized spins are ferromagnetically aligned between the holon and spinon leading to a string or confining 
(linear) potential.  (b) Scheme of the two-hole ground state.  The holons carry solitons with 
opposite charge and are bound together by a magnetic string.  (c) Formation of holon-holon bound 
state from two free holes.  The two spinons with opposite spins and topological charges cancel each other, leaving 
two holons with opposite topological charges attached by the string.  (d)  $C_{hs}(r)$ for a single hole.  The 
arrow indicates the position of the spinon.  (e) $C_{hs}(r)$ for two holes with $\alpha = 1.0$, $J_K = 0.05$, and $L = 100$.  
The reference holon is at $r=1$.  The arrow indicates the position of the second holon.  Each holon carries an ADW as explained 
in the text.  }
\label{Fig1}
\vspace*{-0.5cm}
\end{figure}

We first consider the simplest case of Ising-like exchange interactions ($\alpha=0$)
and a Kondo coupling much smaller than the rest of the terms in ${\tilde H}$. In this situation, 
the exchange $J_{H}$ forces 
the localized spins to be AFM ordered in the ground state. For the $J_K=0$ ground state, each 
hole added to the conduction band carries a soliton or anti-phase domain wall (ADW) for the AFM order 
parameter \cite{Batista00}. For $J_K\neq0$, the single hole quasiparticle becomes a spinon-holon 
bound state [see Fig.\ref{Fig1}(a)] to avoid
an energy increase of order $LJ_K$ \cite{note2}. The single added hole is topologically neutral: 
the holon and the spinon carry
solitons with opposite ``charge'' (kink and anti-kink). We also note that unless the spinon and the 
holon are on the same
site, the ferromagnetic link associated to the spinon increases the magnetic energy by $J/2$. This provides 
an additional attractive force between the spinon and the holon.  In this fully anisotropic limit, the spinon is 
immobile and the holon is localized around it \cite{Smakov07}.

The situation is qualitatively different for the two-hole ground state [see Fig.\ref{Fig1}(b)]. 
The spinons attached to each holon cancel each other (they have opposite spins and topological charges) leading to a bound 
state of two holons attached by a magnetic string. The cancellation of the two 
spinons lowers the magnetic energy by $\sim J$. 
Consequently, we expect the two-holon bound state to be more stable than two independent 
spinon-holon pairs. This statement can be quantified by computing the exact ground 
state energies for one and two holes (this can be done for $\alpha=0$ because 
the spins are not exchanged by ${\tilde H}$). In particular, for $J_K=0.1$, we get a binding 
energy $\Delta_B = E_g(N_h) + E_g(N_h-2) - 2 E_g(N_h-1) 
\simeq -0.25$ for $N_h=2$, where $E_g(N_h)$ is the ground state energy for $N_h$ holes.   The holon-holon pair 
formation is illustrated in Fig.\ref{Fig1}(c).  The mutual cancellation of the two spinons 
leaves the two holons attached by a magnetic string. 

This simple picture for one and two holes 
has been discussed previously in the context of a $t-J$ model in a staggered magnetic field \cite{Bonca92}.
In our case, the staggered magnetic field $h$ is not artificial because it is self-generated  by the 
AFM ordering of the localized spins. 
As it was pointed out in  \cite{Bonca92}, the limit $h \to 0$ ($J_K \to 0$ in our case) is singular:
lim$_{h\to 0} \Delta_B = -J$ while the mean distance between the 
two holes diverges as $h^{-1/3}$ ($J_K^{-1/3}$). This singular behavior is a manifestation of 
the qualitative difference between the single- and two-hole
states: {\it the binding energy $\Delta_B$ and the size of the two-holon bound state, $l_p$, are determined
by two independent energy scales}. While $\Delta_B \sim -J$ for small enough $h$ ($J_K$), $l_p$ only depends 
on $h$ ($J_K$) as long as $J$ is nonzero. In particular, this shows that a negative binding 
energy, $\Delta_B < 0$, does not imply the formation of a two-hole bound state.
The negative value of $\Delta_B$ for $h \to 0$ ($J_K \to 0$) just indicates that a single hole always creates a spinon
[see Fig.\ref{Fig1}(a)] while this is not true for the two-hole state as shown in Figs.\ref{Fig1}(b) and (c). 
An infinitesimal field $h$ ($J_K$) is enough for stabilizing the bound state due to the topological
nature of the two-hole state: each hole carries a soliton and the two solitons become confined for any
finite $h$. This remarkable property makes the pairing robust against the inclusion of a more realistic
longer range Coulomb interaction in $H$. We will see below that this pairing mechanism survives
in the absence of long range AFM order, i.e., when the effect of the localized spins {\it cannot} be replaced by
a staggered mean field $h$ because $\langle {\bf S}_{i\eta} \rangle=0 $. This is a very important qualitative
difference relative to the case considered in \cite{Bonca92}.
\begin{figure}[t]
\vspace{-1.0cm}
\includegraphics[angle=0,width=8cm]{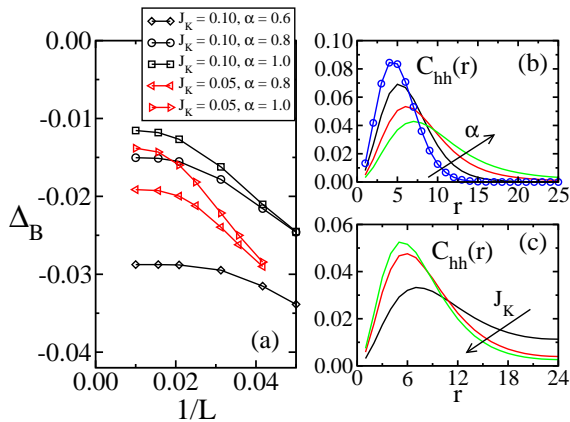}
\vspace*{-0.5cm}
\caption{(Color online) (a) Finite size scaling of $\Delta_B$.  The results indicate robust pairing in the 
thermodynamic limit for for all values of $\alpha$.   
(b) Density-density correlation function $C_{hh}(r)$ for $J_K = 0.1$  and $\alpha = 0.0, 0.6, 0.8, $ and $1.0$.  
$C_{hh}$ clearly confirms the existence of a holon-holon bound state.  Exact results are shown in open circles 
for the Ising limit ($\alpha=0$) and the agreement with DMRG is excellent.  (c) $C_{hh}(r)$ in the fully isotropic 
limit ($\alpha=\beta=1$) for $J_K=0.2, 0.3,$ and $0.35$.  A bound state is formed in this limit.} 
\label{Fig2}
\vspace*{-0.5cm}
\end{figure}

The above picture remains valid away from the Ising limit ( $\alpha > 0$).  However, the hole 
can now  move coherently in one sublattice.   
The process involves the exchange of the two spins on adjacent nearest and next-nearest neighbor sites 
of the hole via the $\alpha J$ term followed by two hoppings in the same direction.   
This coherent motion of the hole leaves the 
magnetic  background unchanged.  For  $0 \le \alpha \le1$,  the holon and spinon form a bound state.  The magnetic structure of this quasiparticle 
is shown schematically in Fig. \ref{Fig1}(a).  
To confirm the above picture, we compute the hole-spin correlation function 
$C_{hs}(r) = \langle S^z_i n^h_{i+1} S^z_{i+r}\rangle$, 
where $n^h_j = 1-(n_{j\uparrow} + n_{j\downarrow})$ is the hole density at site $j$.
Figure \ref{Fig1}(d) shows $C_{hs}(r)$ in the isotropic limit ($\alpha=1$) with one hole 
in the conduction band.  The correlator shows the 
spinon (indicated by the arrow) separated from the holon (at $r=1$) by a finite distance.  
$C_{hs}$ shows clearly that the holon and spinon carry an ADW. 
When a second hole is added to the conduction band, the spinons cancel each other, as explained above, 
leaving the two holons attached to ADW's of opposite sign.  This is also confirmed by the calculation of $C_{hs}$ 
[see Fig. \ref{Fig1}(e)].  The reference holon is at $r=1$ while the average 
position of the second holon is indicated by the arrow (note that this average distance is artificially increased due
to the PBC).  To the left side of the second holon, the spins at even 
numbered sites ($r=2$ for example) are antiparallel to the reference spin; 
whereas to its right, they are parallel (e.g. $r=32$).

In the following we present numerical evidence of pairing and dominant 
superconducting correlations as a function of hole density $\nu=N_h/L$.  
Figure \ref{Fig2}(a) shows the pairing energy $\Delta_B=E_g(2) + E_g(0) - 2 E_g(1)$ in the dilute limit $\nu \to 0$
versus $1/L$ and for different values of $J_K$ and $\alpha$.  The finite size scaling 
of $\Delta_B$ shows robust pairing in the thermodynamic limit for all 
values of $\alpha$.  Figure \ref{Fig2}(b) shows the density-density correlation function 
$C_{hh}(r) = \langle n^h_{i} n^h_{i+r}\rangle$ for $J_K = 0.1$ and 
different values of $\alpha$.  $C_{hh}$ shows a clear maximum at a finite distance confirming 
that the two holons are bound.  As expected, the distance between the two holons increases with $\alpha$.   
In the Ising limit, $\alpha = 0$, the DMRG results are compared to the exact solution and 
the agreement is excellent.  

\begin{figure}[t]
\vspace{-1.1cm}
\includegraphics[angle=0,width=8cm]{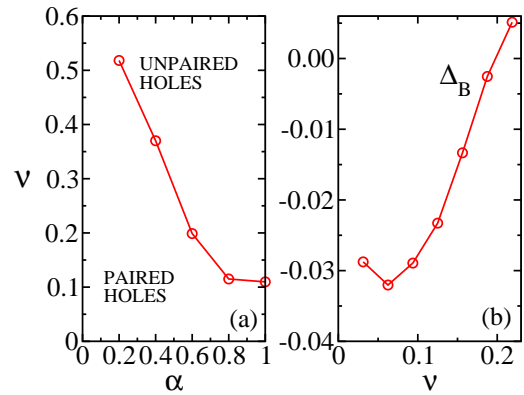}
\vspace{-0.5cm}
\caption{(Color online) (a) $(\alpha,\nu)$ phase diagram for $L=64$ and $J_K=0.1$.  Near the Ising 
limit ($\alpha=0.2$), the pairing survives up to $\nu\approx 0.5$ ($N_h\approx32$).  
For a given $\alpha$, the boundary between the two phases is 
determined from the condition $\Delta_B(\alpha,\nu_c)=0$ with $\nu_c$ being the lowest value of $\nu$ which satisfies this
condition (holes are not bound for  $\nu>\nu_c$).
(b) $\Delta_B(\nu)$ for $\alpha=0.6$.  In the thermodynamic limit ($L\to \infty$), 
$\Delta_B$ should be zero for $\nu>\nu_c$; however 
$\Delta_B(L=64)>0$ due to finite size effects.}
\label{Fig4}
\vspace*{-0.5cm}
\end{figure}

So far we have only considered the limit $\beta=0$ because the pairing mechanism is easier to identify 
in the presence of long-range AFM order. It is natural to ask if the pairing survives in the fully isotropic 
limit ($\alpha=\beta=1$)  relevant for our original Hamiltonian $H$. In the absence of holes, the ground
state of ${\tilde H}$ only exhibits short-range  AFM correlations due to the gap induced by the relevant 
$J_K$ coupling. The $C_{hh}$ correlator for the two-hole ground state shown in Fig.\ref{Fig2}(c) provides
clear evidence of the formation  of  a two-hole bound state. Moreover, the pair size $l_p$ decreases monotonically
with increasing $J_K$ (slope of the confining potential) indicating that the pairing mechanism remains the same.

\begin{figure}[t]
\vspace{-0.8cm}
\hspace*{-0.7cm}
\includegraphics[angle=0,width=10cm]{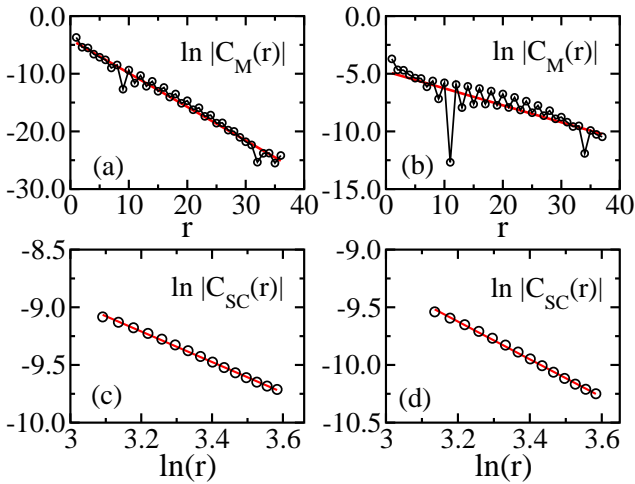}
\vspace{-0.7cm}
\caption{(Color online) Single particle, $C_M$, and pair-pair, $C_{SC}$, correlation functions.  
The circles show the numerical results, and the straight 
lines show the best fit.  (a), (c) $J_K=0.05$, $\alpha=0.2$. (b), (d) $J_K=0.05$, $\alpha=0.6$.  
$C_M(r)$ shows an exponential decay whereas $C_{SC}(r)$ shows a power law decay.  This confirms 
the emergence of dominant superconducting 
correlations.}
\label{Fig3}
\vspace*{-0.5cm}
\end{figure}

For finite hole concentration, a new length scale $1/\nu$, that is the mean distance 
between holes, appears in the problem. We expect the previous analysis of the dilute limit to remain valid as long
as  $l_p \ll 1/\nu$. On the other hand, the pairing should be suppressed 
when the these two lengths become comparable  because the effective interaction between holes is repulsive 
at short distances. This expectation is fully consistent with the numerical results shown 
in Fig.\ref{Fig4}. The boundary between the paired and unpaired regions (see Fig.\ref{Fig4}a) is shifted
to lower hole concentrations when $\alpha$ gets closer to one, i.e., when $l_p$ becomes bigger. 
In particular, Fig.\ref{Fig4}b shows the evolution of the pairing energy $\Delta_B$ as a function of $\nu$
for $\alpha=0.6$. $\Delta_B$ ceases to be negative for a hole concentration close to 20\% 
($\nu_c \sim 0.2$). This implies that if we vary the chemical potential, the states with odd number of holes
are metastable as long as $\nu < \nu_c$. We note
that the critical hole concentration $\nu_c$ remains significantly high ($\nu_c \sim 0.1$) in the limit $\alpha=1$.

To confirm the presence of dominant superconducting correlations, we compare the 
single particle $C_{M}(r) = \sum_{\sigma}\langle \bar c^{\;}_{i+r\sigma} \bar c^\dagger_{i\sigma}\rangle$ 
and pair-pair $C_{SC}(r) = \langle \Delta_{i+r}\Delta^\dagger_{i}\rangle$ correlation functions, 
with $\Delta_i^\dagger = {1\over\sqrt{2}}
(\bar c^\dagger_{i+1\uparrow} \bar c^\dagger_{i\downarrow}- \bar c^\dagger_{i+1\downarrow} \bar c^\dagger_{i\uparrow})$.
Figure \ref{Fig3} shows results obtained for $L = 80$, $N_h = 4$, $J_K = 0.05$ 
and $\alpha = 0.2, 0.6$.  The single particle correlator decays exponentially while the 
pair-pair correlator shows a slower algebraic decay. The exponential decay of the single 
particle correlator is due to the fact that holes are bound in pairs. In the dilute limit, the probability of finding the two
holes separated by a distance $r$ bigger than the pair size $l_p$ (see Fig.\ref{Fig2}b) decreases exponentially in $r$. 
The algebraic decay of $C_{SC}(r)$ is expected for a Luttinger liquid of pairs. The comparison between 
$C_{M}(r)$ and $C_{SC}(r)$ clearly indicates that the ground state has dominant superconducting 
correlations in the regime under consideration.

A few comments are in order. We verified that the pairing ($\Delta_B<0$) persists for smaller 
values of $J_H$ such as $J_H=J_K=0.1$ .
The small $J_H$ regime of ${\tilde H}$ is relevant for describing lanthanide and actinide based compounds
in which localized f-electrons interact via Kondo exchange with electrons in the conduction band. Conduction bands with 
strong $3d$-character are correlated and would provide a natural realization of our ${\tilde H}$. If the $t-J$ conduction band 
of ${\tilde H}$ is replaced by the original Hubbard upper band of $H$, one can study the evolution of our 
pairing mechanism as a function of $U_u/t$ (the standard KL model is recovered for $U_u/t=0$) \cite{Khaled}.

In contrast to pairing mechanisms driven by an attractive short range interaction (like the AFM exchange J), our
mechanism is robust under the inclusion of a more realistic longer range Coulomb interaction. This results from
the fact that $\Delta_B$ and $l_p$ are determined by two independent energy scales ($J$ and $J_K$). In other words,
a big enough  value $l_p$ reduces the effect of longer range Coulomb terms without reducing the value of
$\Delta_B$ (note that $l_p$ and $\Delta_B$ are anticorrelated when the pairing is produced by a short range 
attractive potential).

Finally, our pairing mechanism should also persist for weakly coupled chains (small inter-chain hopping)
due to its topological nature. While single-particle coherent inter-chain hopping  is not possible due to the
soliton that is attached to each hole, a pair can hop coherently between chains because it is topologically neutral 
(soliton-antisoliton bound state). Such coherent pair-hopping should stabilize a superconducting state below a finite critical
temperature. A detailed study of this extension to higher-dimensional systems will be presented elsewhere \cite{Khaled}.

The authors thank E. Dagotto, J. Bonca, S. Trugman, U. Schollw\"ock, G. Martins, and C. B\"usser for helpful discussions.  LANL is supported by the 
U.S. DOE under Contract No. W-7405-ENG-36.


\vspace*{-0.4cm}


\begin{thebibliography}{99}

\vspace*{-0.4cm}

\bibitem{Anderson07} P. W. Anderson, Science {\bf 317}, 1705 (2007).
\bibitem{Dagotto94} E. Dagotto, Rev. Mod. Phys. {\bf 66}, 763 (1994).
\bibitem{Tsunetsugu97} H. Tsunetsugu, {\it et al.}, Rev. Mod. Phys. {\bf 69}, 809 (1997); 
I. P. McCulloch, {\it et al.}, Phys. Rev. B {\bf 65}, 052410 (2002); 
D. J. Garcia, {\it et al.}, Phys. Rev. Lett. {\bf 93}, 177204 (2004); 
J. C. Xavier and E. Dagotto, Phys. Rev. Lett. {\bf 100}, 146403 (2008).
\bibitem{Khaled} K. A. Al-Hassanieh et al, to be publised.
\bibitem{White92} S. R. White, Phys. Rev. Lett. {\bf 69}, 2863 (1992); Phys. Rev. B {\bf 48}, 10345 (1993).
\bibitem{Hallberg06} K. Hallberg, Adv. Phys. {\bf 55}, 477 (2006); U. Schollw\"ock, Rev. Mod. Phys. {\bf 77}, 259 (2005).
\bibitem{note1}To get good accuracy for this model, it is useful to keep a large number of states in the infinite-system DMRG step and to consider each unit cell as one lattice site with $2\times3$ states.
\bibitem{Batista00} C. D. Batista and G. Ortiz, Phys. Rev. Lett. {\bf 85}, 4755 (2000).
\bibitem{note2} This confining mechanism is analogous to the one that binds spinons in pairs
(magnons) when a finite coupling between AFM chains is turned on.
\bibitem{Smakov07} J. Smakov {\it et al.}, Phys. Rev. Lett. {\bf 98}, 266401 (2007).
\bibitem{Bonca92} J. Bonca {\it et al.}, Phys. Rev. Lett. {\bf 69}, 526 (1992); P. Prelovsek {\it et al.}, Phys. Rev. B {\bf 47}, 12224 (1993).

\end{thebibliography}
\end{document}